# An Unpublished Debate Brought to Light: Karl Popper's Enterprise against the Logic of Quantum Mechanics

Flavio Del Santo

Institute for Quantum Optics and Quantum Information (IQOQI), Vienna and
Faculty of Physics, University of Vienna, Austria and
Basic Research Community for Physics (BRCP)

**Abstract**

Karl Popper published, in 1968, a paper that allegedly found a flaw in a very influential article of Birkhoff and von Neumann, which pioneered the field of "quantum logic". Nevertheless, nobody rebutted Popper's criticism in print for several years. This has been called in the historiographical literature an "unsolved historical issue". Although Popper's proposal turned out to be merely based on misinterpretations and was eventually abandoned by the author himself, this paper aims at providing a resolution to such historical open issues. I show that (i) Popper's paper was just the tip of an iceberg of a much vaster campaign conducted by Popper against quantum logic (which encompassed several more unpublished papers that I retrieved); and (ii) that Popper's paper stimulated a heated debate that remained however confined within private correspondence.

1. **Introduction.**

In 1936, Garrett Birkhoff and John von Neumann published a paper aiming at discovering "what logical structure one may hope to find in physical theories which, unlike quantum mechanics, do not conform to classical logic" (Birkhoff and von Neumann, 1936). Such a paper marked the beginning of a novel approach to the axiomatization of quantum mechanics (QM). This approach describes physical systems in terms of "yes-no experiments" and aims at investigating fundamental features of theories through the analysis of the algebraic structures that are compatible with these experiments. The main breakthrough of Birkhoff and von Neumann was to show that (Boolean) classical logic is incompatible with the phenomenology of QM, due to the Heisenberg's uncertainty principle (see section 2).

Karl R. Popper (1902-1994), one of the greatest philosophers of science of modern times, had been working on foundations of quantum mechanics throughout his whole career: since 1934 he had severely criticised the widely accepted "Copenhagen interpretation" of quantum mechanics, advocating objective realism in contraposition to instrumentalism and subjectivism (see Del Santo, 2018; 2019; Freire, 2004, and references thereof). In particular, one of Popper's main critical targets had always been the interpretation of the Heisenberg uncertainty principle that the Copenhagen doctrine provided. Already in his masterpiece *Logik der Forschung* (Popper, 1934), Popper advocated an extreme statistical interpretation of the Heisenberg's uncertainty relations, according to which "these formulae set some lower limits to the *statistical dispersion* or '*scatter*' of the results of sequence of experiments" (Popper, 1967, p. 20), and, as such, they do not say anything about single systems, but only about *ensembles*.

As a matter of fact, around the end of the 1960s, Popper breathed new life into his old critique of the Copenhagen interpretation, also thanks to the considerable help of eminent physicists, the likes of David Bohm, Herman Bondi, Alfred Landé, Wolfgang Yourgrau and Henry Margenau, and he published an influential essay entitled *Quantum Mechanics without the Observer* (Popper, 1967).[1]

Contingently, at the beginning of 1968, Popper became aware that the aforementioned paper by Birkhoff and von Neumann had, by that time, initiated a new subfield of fundamental research in quantum physics, in the spirit of Copenhagen interpretation, known as the "logic of quantum mechanics" (LQM, or "lattice-theoretical approach" to QM, or simply "quantum logic"). His hostile reaction emerges from a letter that he sent to his friend and renown cosmologist, Hermann Bondi:

> I have only heard by accident quite recently that the Birkhoff-von Neumann paper has meanwhile become 'the by now classical work' […], and a school of quantum physicists take it very seriously, and build all sorts of horrible theories on it. (Letter from Popper to Bondi, February 8th, 1968. PA, 96/3)

Popper refers here to the prominent "school" of Joseph Maria Jauch in Geneva, which –starting from the early 1960s– revived Birkhoff and von Neumann's approach and led quantum logic to become rather influential.[2]

Although it is perhaps not too well known to a vast public, a number of works in the literature at least mention that Popper published, in 1968, a critical paper entitled *Birkhoff and von Neumann's Interpretation of Quantum Mechanics*, which appeared in the prestigious journal *Nature* (Popper, 1968). What was, however, hitherto not known is that such a paper was just the tip of the iceberg of what Popper himself referred to as "a greater enterprise directed against the new quantum logic".[3]

The present paper aims at reconstructing Popper's stand on the raising of the impact of LQM, in the context of his broader and decades-long critique of Copenhagen interpretation of QM. By means of new archival sources, I will show that (i) Popper's 1968 paper was just a small part of a much vaster campaign that he conducted against quantum logic. In fact, this endeavour encompassed several more papers that never appeared in print. Moreover, (ii) I will show that Popper's paper stimulated a heated debate that remained however confined within private correspondence.

Starting from early 1968, indeed, Popper devised a series of formal proofs on lattice theory, aiming at disproving the interpretation derived from Birkhoff and von Neumann's pioneering paper, and in particular its justification of the Heisenberg uncertainty principle. Although most of this effort turned out to be based on misconceptions of the original work –and Popper later distanced himself from some of his criticisms of the LQM[4]– I deem this case study remarkable for at least two reasons:

On the one hand, it helps to shed light on the interplay between philosophers and physicists on the ground of foundations of QM, and in particular, on the peculiar role played by Popper at the edge between these communities (see further and cf. Del Santo, 2019). In particular, I will provide new evidence of his forgotten endeavour against the LQM (and the Copenhagen interpretation along with it). Moreover, I will show that there exists proof of at least three unpublished papers authored by Popper (which I could partially retrieve).

On the other hand, I will address what has been called an "unsolved historical issue" (Venezia, 2003), namely that Popper's only published paper on this topic (Popper, 1968), despite appearing in the pages of *Nature*, did not receive any rebuttal in written form until as late as 1974.[5] On the contrary, I will show in detail how the enterprise against the revival of LQM undertaken by Popper led him to enter an intense period of debate with some of its leading proposers (J. Jauch, D. Finkelstein, A. Ramsay, J. Pool). In particular, I retrieved an unpublished rebut of Popper's criticisms by Ramsay and Pool, which was submitted only a few months after Popper's paper appeared in *Nature*, but that was never eventually published.[6] This caused a tremendous controversy (see section 3.2), and I deem it of historiographical relevance to finally reproduce it (as figures in Appendix) after more than 50 years from its conception.[7]

---

1 For a comprehensive reconstruction of Popper's engagement in the research on the foundation of QM and his role in the community of physicists in that period see (Del Santo, 2019).
2 This school comprised scholars such as C. Piron, J. P. Marchand, G. Emch, M. Guenin, B. Misra and others; see e.g. (Jammer, 1974), pp. 351 ff.
3 Letter from Popper to Mario and Marta Bunge in April 1968. (PA, 94/4).
4 In Popper's book on his mature views on the philosophy of quantum theory (Popper, 1982), the editor W. Bartley explicitly states that the parts on "'Birkhoff and von Neumann's Interpretation of Quantum Mechanics' […] contain some points which Popper no longer upholds".

Before getting to the heart of the case we want to investigate, I will contextualize it by briefly recalling what was the status of the research on foundations of QM in that period, as well as Popper's engagement in that debate.

**Intermezzo: Popper's role in the quantum debate**

After an initial period, in the 1920s and 1930s, of heated interpretational debates (among eminent physicists the likes of Bohr, Einstein, Heisenberg, Schrödinger) on the newly established quantum theory, fundamental research experienced a dramatic setback. Largely due to the outbreak of World War II –that also led European physicists to scatter around the world (especially in the US)– the scientific practice drastically changed into a much more pragmatic enterprise. Fundamental questions were mostly substituted by practical problem-solving activities. The subsequent Cold War period did not change things back, and actually fostered pragmatism in physics, leading to the emblematic expression "shut up and calculate!" (see Kaiser, 2011). It was only from the 1960s that a few physicists started breathing new life into fundamental research in QM, voicing the view that the fundamental and the interpretational problems were far from being resolved. In his comprehensive work, O. Freire Jr. calls this generation of physicists "the quantum dissidents" (Freire, 2014). Among them stand the names of J. Bell, H. Everett, D. Bohm, J.-P- Vigier, F. Selleri (see also Baracca, Bergia and Del Santo, 2016). The reason to call them dissidents is twofold: on the one hand, they fought against their *Zeitgeist* to bring back foundations of QM into the focus of physics proper (as opposed to regarding them as "philosophical trivialities"); on the other hand, their main goal was to demolish the hegemony of the Copenhagen interpretation, which, also due to its instrumental approach, became the way of presenting quantum theory without introducing further philosophical issues. On the contrary, the mentioned dissidents strove for bringing back (some forms of) realism into quantum physics, which had been challenged by the Copenhagen interpretation. It ought to be recalled that John Bell put forward the theorem that bears his name in 1964, but that result went essentially unnoticed for over a decade. Today Bell's paper (Bell, 1964) has some 13300 citations and it lays at the basis of the fortunate novel fields of quantum information theory, quantum computation and quantum cryptography. Yet, such was the status of theoretical quantum physics in the 1960s, a formal tool that allowed scientists to make calculations.

Karl Popper, who had been already involved in the foundations of quantum mechanics in the 1930s –when he proposed a mistaken *Gedankenexperiment* that however brought him into discussions with Bohr and Einstein– came back to such problems at the beginning of the 1950s. At that time he devised his propensity interpretation of probability that was eventually to have a long-lasting influence on the physics community. However, as recently reconstructed in (Del Santo, 2019), Popper was throughout all of the 1950s mainly active into the circles of philosophers of science and his resonance in the physics community was close to nothing, even among those dissidents who were already rethinking the foundations of QM. It is interesting to notice that some of Popper's philosophically-inclined acquaintances among the physicists –such as Herman Bondi, Alfred Landé and David Bohm– considered him, in the 1950s, as a reference point to ask for assistance and advice to publish their ideas into journals on the philosophy of science. A major turning point happened in the mid-1960s –mainly thanks to the promotion of Wolfgang Yourgrau, Henry Margenau and Mario Bunge– and gave Popper the visibility to get into the sphere of influence of the quantum dissidents. From that time on Popper's role changed tremendously, and it can be claimed that he became a fully-fledged quantum dissident, in the sense of being active and influential (i.e., participating in their meetings, publishing in their specialized journals, etc.) in the community of physicists concerned with the foundations of quantum mechanics. His international fame as a philosopher and his deep conviction in scientific realism led him to be a central node in the network of these physicists, and he would never lose this status until the end of his days (see Del Santo, 2018). Popper's critique of LQM is thus to be considered, together with his paper *Quantum Mechanics without the Observer* (Popper, 1967), as the main activity that gave him the necessary momentum to cross the disciplinary boundary of philosophy and enter the fight on the foundations of QM on the same ground of the professional physicists. Despite most of his original contributions to quantum physics have turned out to be faulty, his long-standing bridging role between physics and philosophy is to be considered a non-negligible factor in bringing foundations of QM back into proper physics.

---

[5] The only exception is of a succinct footnote in (Jauch and Piron, 1970), see section 3.2.
[6] The existence of this paper is also acknowledged by (Jammer, 1974), and by (Dalla Chiara and Giuntini, 2006) as the only rebuttal of Popper's criticism intended for publication. However, none of these authors ever had the opportunity to get a copy of the paper, and Ramsay and Pool themselves do not have retained a copy of their note to the author on May 15th and September 7th, 2017).
[7] In order to present these results, I have conducted research at Popper's Archive (PA in the references) in Klagenfurt, Austria. The reproduced note was enclosed in a letter from Pool to Jauch on February 19th, 1969, and forwarded to Popper (PA 96/18).

## 2. The Logic of Quantum Mechanics

In order to introduce the historical case, I shall start by giving a short overview of what LQM is about (in this section), and sketching Popper's arguments against it (in section 3.1).[8]

In formal mathematical logic, an algebra ($L$, $\vee$, $\wedge$) on a set $L$ of binary variables is Boolean with respect to the connectives conjunction ($\vee$) and disjunction ($\wedge$), if the following properties hold: commutativity, associativity, existence of the absorbing element, idempotency, existence of maximum and minimum, existence of the complement and distributivity. This latter property states that, given elements x, y, z $\in$ L, x $\vee$ (y $\wedge$ z) = (x $\vee$ y) $\wedge$ (x $\vee$ z).

*Propositional calculus* links these formal properties of mathematical logic to the observables of physical systems, namely it deals with *empirical propositions* and what structure of the logical connectives that relates them is compatible with the observed phenomenology. In this context, an empirical proposition is defined as a binary, decidable statement about a physical system, also called a "yes-no experiment": "these are observations which permit only *one* of two alternatives" (Jauch, 1968). To illustrate this, let us start with classical physics. In classical (Hamiltonian formulation of) physics, the state of a physical system moving in the three-dimensional space is characterized by a point in the (position-momentum) phase-space $\sim \mathbb{R}^6$. An observable $\Omega$ is a real-valued function on this phase-space, such as position, momentum, or the total energy. Given this framework, an empirical proposition is a binary statement about the value of a certain observable in the following form: "The measured value of the observable $\Omega$ lies in the interval $[x, y]$, with $x, y \in \mathbb{R}$". Clearly, the answer is either "yes" or "no". Note that any physical quantity can then be reduced to a collection of empirical propositions (possibly infinite, in the case of continuous variables or unbounded degrees of freedom), dividing the "observational" domain into intervals. These empirical propositions can then be combined using logical connectives and propositional calculus investigates what structures are compatible with the theory under consideration.

---

[8] It is beyond the scope of the present paper to comment in detail the formal aspects of Popper's critiques and the replies he received, which turn out to be somewhat technical. It would be however desirable that this admittedly preliminary reconstruction stimulate more scholars to delve into the details of the controversy here presented and thoroughly analyse the arguments expounded in these unpublished manuscripts whose ubications are here disclosed.

In particular, in his famous *Foundations of Quantum Theory* –the first comprehensive manual of propositional calculus– Jauch shows that the propositions of a physical system (both in classical and quantum physics) form a complete, orthocomplemented lattice.[9] He then clarifies the distinction between formal logic and the calculus of empirical propositions:

> the calculus introduced here [i.e. propositional calculus] has an entirely different meaning from the analogous calculus used in formal logic. Our calculus is the formalization of a set of empirical relations […]. It expresses an objectively given property of the physical system. […] The calculus of formal logic, on the other hand, is obtained by making an analysis of the meaning of propositions. It is true under all circumstances and even tautologically so. […] It turns out, however, that if viewed as abstract structures, they have a great deal in common." (Jauch, 1968, p. 77)

It is, therefore, possible to directly exploit results and theorems of formal logic and adapt them to describe the structure of physical systems. In particular, it is possible to show that the structure of the possible compositions (through logical connectives) of empirical proposition in classical physics form a Boolean algebra, as defined above.

On the contrary, the momentous paper by Birkhoff and von Neumann of 1936 proved that for compositions of quantum empirical propositions the law of distributivity fails. In such a way, the structure of experimental propositions would still be an orthocomplemented lattice, though not a Boolean one. The reason for weakening the classical logical structure was rooted in (the orthodox interpretation of) Heisenberg's uncertainty relations, which imply a violation of the principle of the excluded middle for experimental propositions. Indeed, in quantum mechanics, even for maximal pieces of information about a system (pure states), the full information about one observable (e.g. "spin along x-direction is up"), does not allow to decide the truth state of a second non-commuting observable (e.g. to decide on the truth of the experimental proposition: "spin along the y-axis is down"). Birkhoff and von Neumann, indeed, noticed that the distributive law "is a […] logical consequence of the compatibility of the observers." (Birkhoff and von Neumann, 1936). Under this evidence, they proposed to substitute the distributive law with the weaker modular law, i.e., $((x \wedge z) \vee y) \wedge z = (x \wedge z) \vee (y \wedge z)$.[10]

### 3. Popper's rebut of Birkhoff and von Neumann

Popper had a formal education in logic (and some in mathematics and physics) and started working on problems of lattice theory as early as 1938.[11] As a consequence of the popularity that LQM was gathering – an approach firmly in the spirit of the operational Copenhagen interpretation, since it elevates Heisenberg's uncertainty (in its standard interpretation) to *the* fundamental principle – Popper was called to vindicate his standpoint. He thus published his mentioned paper in *Nature* (Popper, 1968), wherein he claimed that Birkhoff and von Neumann's proposal was untenable, due to an alleged mathematical mistake made by the two preeminent mathematicians (Popper's paper is a formal paper on lattice theory). In Popper's words: "no more than a simple slip – one of those slips which, once in a lifetime, may happen even to the greatest mathematician" (Popper, 1968).

---

[9] A lattice is a set with an operation of partial order $\leq$, and whose pairs of elements each have a unique supremum and infimum. Given the "tautological element" (i.e. the total subspace, denoted by **1**) and the "absurd element" (i.e. the null subset, denoted by **0**), for every element $x$ of the lattice $L$, $\mathbf{0} \leq x \leq \mathbf{1}$. A lattice is said to be *complete* if every subset has both supremum and infimum. Two elements $x, y \in L$ are said to be the complement of each other if $x \wedge y = \mathbf{0}$ and $x \vee y = \mathbf{1}$. A lattice is called *orthocomplemented* if it is provided with an operation of orthocomplementation ('), such that for all $x, y \in L$, $x \leq y \Rightarrow y' \leq x'$ and $(x')' = x$.
[10] It ought to be remarked that further proposals developed by the school of Jauch, especially by his pupil Constantine Piron, showed that modularity may still be a too strong condition and thus the logical structure of QM should be further weakened.

Popper's proposal is based on three arguments of rather different nature: an algebraic one, a probabilistic one and a *Gedankenexperiment*. Popper's first argument goes like this: Birkhoff himself proved in his book *Lattice Theory*, in 1940, that every orthocomplemented lattice is a Boolean algebra provided it is uniquely complemented. Popper notices that orthocomplementation is usually taken to be unique, and that "Birkhoff and von Neumann constantly speak of 'the' complement and 'the' operation of complementation", and thus he infers that the complement was implicitly assumed to be unique by the authors. Thus, Popper concludes that Birkhoff and von Neumann's paper "culminates in a proposal which clashes with each of a number of assumptions made by the authors" (Popper 1968). While this algebraic argument is mostly based on semantic issues, Popper's second probabilistic argument is subtler. Popper is here concerned with the definition of quantum probability as a (bounded and additive) real-valued measure function $m$ on the lattice $L$, with the following rules: (R1) if $x = y$, then $m(x) = m(y)$; (R2) if for every $m$, $m(x) = m(y)$, then $x = y$; (R3) for every $m$ there exist two real numbers $k_0$ and $k_1$, such that $k_0 = m(\mathbf{0})$ and $k_1 = m(\mathbf{1})$; (R4) for every $m$, $m(x) + m(y) = m(x \wedge y) + m(x \vee y)$.[12] Based on these axioms, Popper then proves a series of theorems (all formally correct) which show that $L$ has unique complement and thus is a Boolean algebra, allegedly proving Birkhoff and von Neumann wrong. However, all of Popper's later critics (including Scheibe, 1974; Venezia 2003; Dalla Chiara and Giuntini, 2006) agree that rules (R1)-(R4) represent a strengthening of the original assumptions of Birkhoff and von Neumann, and do not translate correctly their original proposal. The aim of the authors was, in fact, to avoid characterizing any probabilities at all, but merely resorting to a dimension-function $m$. Scheibe noticed that "R4 combined with R2 leads to a very strong axiom system of probability" (Scheibe, 1974), and Birkhoff and von Neumann were fully aware that this would have inevitably lead to the collapse of their proposal back to a Boolean algebra. Popper will eventually accept this point thanks to the interaction with S. Kocher in 1974 (see section 3.2).

At last, Popper questions the validity of a *Gedankenexperiment* devised by Birkhoff and von Neumann as a physical justification to challenge the law of distributivity. Denote by $x$ the experimental observation of a quantum wave packet on one side of a plane and by $x'$ the observation of the same packet on the other side of the plane. Call $y$ the observation of the wave packet in state symmetric with respect to the plane, then Birkhoff and von Neumann maintain:
$$y \wedge (x \vee x') = y \wedge \mathbf{1} = y > \mathbf{0} = x \wedge y = x' \wedge y = (x \wedge y) \vee (x' \wedge y).$$
This expression shows that the equality between the first and the last terms does not hold, and thus the distributivity law breaks down. However, Popper maintains that there is nothing quantum in this example and that "we may substitute an elephant [...] for the wave packet". In fact, he claims, $x'$ is not to be interpreted as the property of being "in the other side of the plane", but rather as "not being on the one side". In such a way, $y = x \wedge y$ and the equality is restored (and as such also distributivity). To my knowledge, nobody successfully rebutted this part of Popper's criticism, because, in Scheibe's worlds, "it has to be admitted that the passage to which Popper refers in this connection is very ambiguous." (Scheibe, 1974). The *Gedankenexperiment* that Popper criticize is, in fact, perhaps the weakest part of the original Birkhoff and von Neumann's proposal, for it is more connected to the physical interpretation of the formalism than to any possible mistake made by the two mathematicians.

---

[11] Popper himself recalled this, in 1968, in the introduction of a preprint of a never-published paper (attached to a letter to Mario and Marta Bunge on April 9th, 1968. PA, 94/4). However, the first article published by Popper on lattice theory of which I am aware of is (Popper, 1947).

[12] In (Popper, 1968), rule (R4) reads instead: for every $m$, $m(x) + m(y) = m(\mathbf{1}) + m(x \vee y)$, due to a misprint. Indeed, Popper found the mistake only after the paper was published and sent the correction to the misprint to several physicists and philosophers (see section 3.2) on October 8th, 1968, (PA, 96/1). It ought to be remarked that both Scheibe (1974) and Venezia (2003) noticed the mistake, but the latter corrected it the wrong way, namely substituting $m(\mathbf{0})$ for $m(\mathbf{1})$.

The fact that Popper's thesis is completely self-consistent has hardly been questioned, but it is the opinion of most of his commentators (see e.g. Venezia, 2003 and Dalla Chiara and Giuntini, 2006) that Popper misinterpreted Birkhoff and von Neumann, who have, in fact, never intended to assume the unicity of the complement. Also Scheibe, in his critical reply to Popper, affirmed "that two mathematical arguments of Popper, valid as they are, rest on interpretative premises completely without foundation in the position actually held by Birkhoff and von Neumann" (Scheibe, 1974), and he pointed out that they could not have assumed a unique complement, since already in "von Neumann's lectures [...] given in the academic year 1935-6: There irreducibility is characterised as stating that apart from the two neutral elements of a lattice no element has a unique complement". It has also been stressed in (Venezia, 2003) and (Jammer, 1974) that Birkhoff and von Neumann's paper is definitely not an example of clarity and actually contains a number of ambiguities and it is indeed necessary to resort to other works of the same authors (as also Scheibe did), or to successive developments by others, to ultimately prove Popper wrong.

In any case, we are here concerned with a case that, from the genuinely historiographical point of view, is a really peculiar one. In fact, it has been stressed that Popper's paper against LQM "represents a real issue of history of science" because "one would have expected a strong reaction by the upholders of this approach." (Venezia, 2003). However, Popper's refutation of the LQM was rebutted only six years later by the already mentioned paper of Erhard Scheibe, *Popper and Quantum Logic* (Scheibe, 1974). Venezia rightly pointed out that, even if Popper would have been trivially wrong –which he does not think and neither does the present author[13] –"why not to answer publicly and demolish his thesis? It should have been a scientific duty, and, given the high profile of the author, anyone would have gained popularity." (Venezia, 2003). Although this historical episode has gone mostly unnoticed, Jammer's excellent book *The Philosophy of Quantum Mechanics* documents the existence of a paper written by Alam Ramsay and James C. T. Pool (sent to *Nature* in October 1968) in order to vindicate the LQM and of a counter-reply by Popper (sent to *Nature* in February 1969). Jammer anyway concludes that "due to accidental but never fully clarified circumstances none of these papers, although obviously written for publication, has ever appeared in print." (Jammer, 1974, pp. 353). Jammer's reconstruction (partly based on the correspondence he had with Popper at that time) is factual and clean-cut, yet it does not tell the whole story. By means of Popper's correspondence and notes, it is now possible for the first time to give this unsolved historical incident a resolution. Two main new elements emerge: (1) Popper's critique of the LQM was not publicly rebutted but it stimulated an impassionate debate, which has never appeared in print (as I shall show, Ramsay and Pool's note became a substantial part of this debate). Moreover, (2) Popper's paper in *Nature* was only a minor part of a vaster campaign that Popper conducted against the LQM in those years.

---

[13] Popper was surely wrong in maintaining that the orthocomplemented modular lattice of Birkhoff and von Neumann was uniquely complemented. However, the lack of clarity in Birkhoff and von Neumann's paper played a major role in Popper's misinterpretation, which is anyway completely self-consistent. Moreover, I think, with Venezia, that Popper was right when he stated that Birkhoff and von Neumann's proposal was missing a clear physical interpretation in terms of the uncertainty principle. Venezia also claimed that "it exists an objective difficulty in the literature to rebut Popper about his probabilistic interpretation of" Birkhoff and von Neumann. For all these reasons I think that one cannot affirm that Popper was *trivially* wrong.

### *3.1 A greater enterprise directed against the new quantum logic*

As it was argued before, Popper's motivation for such an overdue intervention was rooted in the revival that the LQM was experiencing in the 1960s. Popper explicitly explained this to Bondi, in the letter that inaugurated his activities conducted against the LQM:

> I am writing to ask for your advice. […] It so happens that I had for many years a proof that [Birkhoff and von Neumann's] proposal is inconsistent. […] I have only heard by accident quite recently that the Birkhoff-von Neumann paper has meanwhile become 'the by now classical work' (Jammer, The Conceptual Development of Quantum Mechanics, p. 376) [(Jammer, 1966)], and a school of quantum physicists take it very seriously, and build all sorts of horrible theories on it. So I have written two papers. […] The two papers originally formed one. I cut them into two because I thought a shorter paper may be published quickly. What worries me is the first paper. […] Can you help?" (Popper to Bondi on February 8th, 1968. PA, 96/3).

It appears clear that Popper was seriously intentioned to put an end to the development of the LQM and wanted to do it as quickly as possible. Bondi got immediately down to work: firstly, he suggested to Popper to contact David Bohm, since the latter had recently published, together with Jeffrey Bub, a paper against Jauch and Piron's refutation of hidden variables in quantum mechanics (Bohm and Bub, 1966). Furthermore, he advised Popper to send one of the papers as a letter to the Editor of *Nature*, whilst Bondi himself would have submitted the second paper for publication, on Popper's behalf, to the *Royal Society* (of which Bondi was a fellow).

Indeed, on February 27th, 1968, Popper wrote to the editor of *Nature*, John Maddox, urging the publication of *A Note on Birkhoff and von Neumann's Interpretation of Quantum Mechanics*:

> The enclosed letter to the editor of Nature was written on the suggestion of Professor Hermann Bondi: it is an extract from a longish paper which, I fear, is far too long and has too many formulae for Nature. I am most anxious to have this letter published as soon as possible. (The topic is one on which a whole crowd of people are working at present). (PA, 94/9).

This paper, which was eventually published as a general article (Popper, 1968) and not as a letter to the Editor,[14] was a summary of only one of the nine sections of which the long original paper consisted. Popper asserted that "in fact, the letter to Nature overlaps not even marginally with the paper" he sent (through Bondi) for publication in the proceedings of the *Royal Society*. (Letter from Popper to Bondi on March 5th, 1968. PA, 96/3).

Popper also called Bohm at the telephone, who sent him an offprint of (Bohm and Bub, 1966). However, Popper maintained afterwards:

> this does not in the least overlap with my approach: both my paper and my letter to Nature are strictly formal refutations based on Birkhoff's own book Lattice Theory (which neither Piron nor

---

[14] The paper was accepted on July 15th, 1968, and the editorial decision of moving to the general articles was also communicated (letter from *Nature* to Popper. PA, 96/23).

Jauch (nor, I am afraid, Bohm) utilize). […] My paper is so formal that it can be checked by a computer (PA, 96/3).

At the same time, Popper communicated to Bondi's secretary the second, and much longer, paper,[15] *On a Conjecture by Garrett Birkhoff and John von Neumann*, which was submitted to the *Royal Society* by Bondi in the following days (letter from Bondi to Popper on March 9th, 1968. PA, 96/3). Popper also advised Bondi on the possible choice of the referees: "I cannot suggest someone, but I certainly would not want a philosopher or even a logician as a referee. […] the best would be a mathematician (algebra? Topology?) or a physicist, with some common sense, and not dependent on the Copenhagen interpretation. (I fear this is asking a great deal)". (Letter from Popper to Bondi on March 5th, 1968. PA, 96/3). However, the Executive Secretary of the *Royal Society* replied to Bondi, on May 21st, 1968, that Popper's paper had been rejected because the referee did not report in favour of publication. The review, based on arguments very similar to those that Scheibe will have used some years later, wrote that the paper "is internally consistent but rests upon a misinterpretation of the work of Birkhoff and von Neumann" (a copy of the referee report is available in PA, 96/3). Bondi forwarded the decision to Popper, commenting that "the Royal Society have [*sic*] indeed been stupid". Despite Popper submitted a detailed reply to the referee (Popper to Ms. G. L. Browne, Bondi's Secretary, on June 17th, 1968. PA, 96/3), on the very same day he also privately wrote to Bondi to express his concern about the possibility of the referee changing his mind: "the criticisms of the referee can indeed easily shown to be false […]. However, I feel that the referee is so irresponsible that he is unlikely to give in." (Popper to Bondi on June 17th, 1968. PA, 96/3). Popper's uneasiness led him, a month later, to write again to Bondi, stating: "I feel I have waited long enough, and I have decided to withdraw my paper from the R. S." (Popper to Bondi on July 17th, 1968. PA, 96/3).

Meanwhile, Popper had orchestrated more. While the two papers submitted to *Nature* and to the *Royal Society* were undergoing the review process, Popper had prepared another manuscript of intermediate length "containing a mathematical criticism of the famous paper by Garrett Birkhoff and John von Neumann" (PA, 96/2). Popper submitted this new paper for publication to the *International Journal of Theoretical Physics* (on July 9th, 1968) and to *Il Nuovo Cimento* (on July 14th, 1968). This paper too, however, was eventually never published.[16]

In addition, as early as April 1968, Popper wrote to Mario Bunge and his wife Marta, who at that time was a young researcher in mathematics. Popper contacted her to ask for feedback about yet another paper he wrote "in Marta's field" (PA, 94/4). In that letter, Popper explains that "it contains a proof that all lattices with unique complements are Boolean Algebras. This proof flies right in the face of a famous result by R. P. Dilworth, a result accepted by Birkhoff and Mac Lane" (April 9th, 1968, Popper to Mario and Marta Bunge. PA, 94/4). The eight-page long manuscript, enclosed in the letter, aimed at disproving the result of Dilworth and thus was part of Popper's enterprise to undermine LQM. However –as also accepted by Popper in his paper in *Nature* (Popper, 1968)– only specific types of lattices (e.g. orthocomplemented or modular) are Boolean algebras provided they are uniquely complemented. Therefore, Popper's proposal was this time undoubtedly flawed. At any rate, Popper sent, on the same day, copies of his proposal to the major authorities in the field of lattice theory – some of the most influential living mathematicians of that time – who Popper was severely criticising: Stephen C. Kleen, (PA, 94/7), Saunders Mac Lane (PA, 94/8), Garret Birkhoff (PA, 277/10) and Robert P. Dilworth himself (PA, 94/5). With this, Popper aimed at throwing himself into an open debate over the mathematical foundations of quantum logic. However, mathematicians have completely ignored Popper's attack and even Birkhoff who was –together with von Neumann, who had however passed away in 1957– the main target of the whole enterprise against LQM, never answered any of the letters that Popper sent him. As we shall see in the next section, it was only the publication of the paper in *Nature* that triggered an intense debate with the school of physicists concerned with LQM.

---

[15] From the review of the referee of the Royal Society (PA, 96/3), one can infer that the paper was at least 21 pages long. I could, however, find a copy with only the first 18 pages in (PA, 465/7).

[16] About this paper we only have indirect evidence thanks to Popper's letters, but no copies seem to have survived. However, it is clear this is a different paper from the previous two, because Popper states, in the letters to the editors: "a paper of somewhat less than 5000 words" (PA, 96/2); whereas the paper in Nature was around 2800 words long and the one submitted to the Royal Society much longer (at least 21 pages). It was Popper himself who withdrew the submission of his paper from the *International Journal of Theoretical Physics*; on September 14th, 1968, he wrote to the editor John Yates stating that "although this paper was much more detailed, its scope did not go beyond the outline that [he] gave on Nature", (PA, 311/40).

It is now evident that what has gone down in history as an isolated incident was actually a minor evidence of a number of activities that Popper was organising, urging to stop the quantum logic approach. The table below lists all the manuscripts that Popper wrote against the LQM, and the additional documents related to them, none of which ever appeared in print with the only exception of (Popper, 1968).

| **Papers authored by Popper on the LQM** | | | |
|---|---|---|---|
| Title | Description | Submitted to | Status |
| *Birkhoff and von Neumann's interpretation of Quantum Mechanics* | ~2800 words; on an alleged mathematical error in (Birkhoff and von Neumann, 1936) | *Nature* (27/02/1968) accepted: 15/07/1968 published: 17/08/1968 | Published (Popper, 1968) |
| *On a Conjecture by Garrett Birkhoff and John von Neumann* | 21 pages; on an alleged mathematical error in (Birkhoff and von Neumann, 1936) | *Royal Society* (through H. Bondi, 09/03/1968); refuted by the Referee (see ancillary documents) | Unpublished; (Partially?) retrieved (PA, 465/7) |
| N/A | ~5000 words; on an alleged mathematical error in (Birkhoff and von Neumann, 1936) | *International Journal of Theoretical Physics* (09/07/1968); *Il Nuovo Cimento* (14/07/1968) | Unpublished; lost |
| *The Distributivity of Lattices with Unique Complements* | 8 pages; on an alleged proof that all lattices with unique complements are Boolean Algebras; against R. P. Dilworth's results | N/A | Unpublished; Retrieved (PA, 94/4) |
| *Reply to Ramsay and Pool* | 6 pages; reply to the paper "Remarks on a Paper by Karl R. Popper" by A. Ramsay and J. C. T. Pool (see supplementary documents) | *Nature* (June 1969) | Unpublished; Retrieved (PA, 96/23) |
| **Supplementary documents** | | | | |
| Author | Title | Description | Submitted to | Status |
| Referee of the *Royal Society* | Referee's report | 1 page | *Royal Society* | Not intended for publication; Retrieved (PA, 96/3) |

| K. R. Popper | Reply to the Referee's Report | 1 page | *Royal Society* | Not intended for publication; Retrieved (PA, 96/3) |
| A. Ramsay an J. C. T. Pool | Remarks on a Paper by Karl R. Popper | 3 pages | *Nature* (16/10/1968) | Unpublished; Retrieved (PA, 96/18) [reproduced in Appendix] |

### *3.2 The unpublished debate between Popper and the school of the LQM*

Although only one of a number of papers written by Popper was published, this appeared on the highly read journal *Nature*, and hence it did not go unnoticed, contrarily to what standard historiographic reconstruction have thus far reported. Additionally, it was Popper himself who started a campaign of popularisation of his critique of the LQM. He thus promoted his activities in the course of conferences and meetings. Popper gave, for instance, a talk on this topic in Vienna in late 1968 (recalled by the mathematician and philosopher Richard Montague and by Mario Bunge).[17] Moreover, Popper sent an offprint of his *Nature* paper to a multitude of distinguished academics, between logicians, mathematicians, theoretical physicists and philosophers of science. Just to mention some notable ones: K. Menge, W. Kneale, A. Robinson, R. Carnap, K. Gödel, V. Kraft, H. Hermes, A. Mostowski, M. H. Stone, M. Bunge, G. Bergmann, G. Birkhoff, R. P. Dilworth, M. Black, W. Craig, J. M. Jauch, A. Landé, R. Harropp, A. Church, R. Lyndon, G. Schütte, G. Müller.[18] Moreover, on October 8th, 1968, Popper sent notification of a misprint in his paper (Popper, 1968) to some other scholars, who clearly already had received a copy of the paper; among them the already mentioned Yourgrau, Bondi, Bohm, Piron, as well as C. W. Kilmister, M. Jammer, C. F. von Weizsäcker, D. Miller, H. J. Groenewald.[19]

This session aims precisely at reconstructing the hitherto unknown resonance that this campaign against LQM has had. Given that the debate that followed Popper's claims against LQM was confined to private correspondence only, I have deliberately chosen to present this section in the form of almost a commented dialogue between Popper and his critics, leaving vast room for excerpts from correspondences.

As already recalled, it was particularly Jauch, who revived the interest towards Birkhoff and von Neumann's proposal of a novel logic for quantum mechanics. Joseph Maria Jauch (1914-1974) obtained his diploma in theoretical physics in Zurich under the supervision of Wolfgang Pauli, of whom he later became an assistant. After various research experiences in prestigious universities in the US and at the CERN (Geneva), in 1960 Jauch was appointed full professor at the University of Geneva, where he became the director of the Institute for Theoretical Physics. Jauch was a fine mathematical physicist, and he devoted a great deal of attention to foundational problems of quantum theory, merging rigorous axiomatic approaches to quantum mechanics (mostly from von Neumann's tradition) with a genuine interest in philosophical issues. He was therefore among the first physicists, in the early 1960s, to revive the interest towards foundations, criticising "the pragmatic tendency of modern research [which] has often obscured the difference between *knowing the usage of a language* and *understanding the meaning of its concepts.*" (Jauch, 1968). In Geneva, Jauch established a prominent school of theoretical physicists whose research activities were often devoted to foundations of QM. However, Jauch's views were far from the ones of most of those *dissidents* who actively opposed pragmatism advocating a realist interpretation of QM, such as H. Everett, B. de Witt, D. Bohm, J. Bell, J.-P. Vigier, F. Selleri, etc. (see e.g. Freire, 2014 and Baracca, Bergia and Del Santo, 2016). As already mentioned, also Popper had supported, since as early as 1934, a strong realist interpretation of QM, against the *subjectivist* Copenhagen interpretation (Del Santo, 2018; 2019). On the contrary, Jauch not only was aligned with Copenhagen but was also a positivist of the kind that Popper abhorred, still supporting inductionism:[20] "Empirical truth […] is synthetic truth. The general physical laws are arrived at by *induction* from observed facts" (Jauch, 1968, p. 70).

---

[17] Letter from Montague to Popper on November 1st, 1968 (PA, 96/21) and Letter from Bunge to Popper September 17th, 1968 (PA, 280/26).
[18] A long well-organised list of names with their addresses and affiliations is attached to a note dated August 9th, 1968 in (PA, 94/1).
[19] These names are reported at the bottom of a note dated October 8th, 1968, which contains the correction to the misprint (PA, 96/1).

Coming back to LQM, early in 1968, Popper informed Jauch that he was about to publish a rebut of Birkhoff and von Neumann's thesis. Jauch replied on February 2nd, 1968: "You refer to a refutation of the Birkhoff-von Neumann paper on merely formal grounds which would be of the utmost interest to me." (Jauch to Popper. PA, 96/18). On August 17th, Popper's paper appeared on the pages of *Nature*, and it did not go unnoticed. Jauch contacted Popper on October 30th, 1968, starting his counteroffensive with a somewhat aggressive approach:

> "it is clear that your interpretation of the paper by Birkhoff and von Neumann disagrees with that of all the workers in the field (such as Piron, Ludwig, Mackey, Varaderajan, Pool, Misra, Finkelstein, and many others), all of whom are in essential agreement how to interpret the propositional calculus of quantum theory […]). Why do you insist in refuting your own private interpretation with an argument which is essentially trivial […]? I feel therefore, your article should be corrected and the best solution would be if you were to do so yourself […] and spare me the disagreeable task of doing it myself in a note to Nature". (PA, 96/18)

And quite naturally, Popper replied rejecting the principle of authority that the letter of Jauch seemed to invoke:

> I shall of course gladly accept any criticism of my paper in Nature and, if this criticism turns out to be valid, I shall gladly write to Nature […].
> 
> You write that my interpretation of the paper by Birkhoff and von Neumann disagrees with a number of authorities. Precisely: this was the reason why I wrote it. (Just as my paper 'Quantum Mechanics without "the Observer" has only one raison d'être: that it disagrees with a number of authorities.)
> 
> […] Your remarks about myself are certainly irrelevant to the problem, especially since I am very ready to make any correction to my paper as soon as I have received some indications as to what I should correct (I cannot very well write to Nature saying that I must be wrong because you and some other authorities say so.) (Popper to Jauch on November 14th, 1968. PA, 96/18)

Jauch replied again urging Popper to consider the modern developments of LQM and not only the original work:

> I am so sorry that we don't seem to communicate very well. You are correct in pointing out some inaccuracies in my language. […] The point at issue is the interpretation of the paper by Birkhoff and von Neumann. If you wish to find out what the authors meant, […] you could of course find

---

[20] Arguably, Popper's most notable contribution to the philosophy of science is his refutation of inductivism as the basis of the scientific method, replaced by his most famous doctrine of *falsificationism*.

> out by contacting Professor G. Birkhoff. I know that he considers your interpretation not to conform to their intention.
>
> If it is a question of interpreting some ambiguities in their text (and there are some, indeed), then the proper way for you to proceed is to study how others who have thought about quantum mechanics have in fact used their ideas and made them more precise […]. (Jauch to Popper on November 22$^{nd}$, 1968. PA 96/18).

Jauch continued stressing Popper's misconception but dismissing it as trivial more than pointing out any concrete criticism. At the same time, he informed Popper that Ramsay and Pool had written a rebut intended for publication: "I have just learned that Professor Pool and Ramsay have sent a reply to Nature which points out your errors and which you will no doubt see in due time." (PA, 96/18).

In fact, the American physicists Arlan Ramsay and James C. T. Pool had sent already on October 16$^{th}$, a critical letter to *Nature*, entitled *Remarks on a Paper by Karl R. Popper*. However, they apparently had contacted neither Popper nor Jauch before the submission. It was the latter who eventually informed Popper, who replied:

> I am very glad to hear that there will be a letter to Nature by Professors Pool and Ramsay. Unfortunately, Nature usually takes quite a long time before forwarding a reply to the original contributor: thus I should be very glad if Professors Pool and Ramsay would send me a copy.
>
> I feel that it is rather difficult for me to write to Professor Garrett Birkhoff as I sent him copies of my letter to Nature long ago without receiving an acknowledgement.
>
> You speak of an ambiguity in the text of Birkhoff and von Neumann's paper and I feel that if this discussion could clear up the ambiguity something worthwhile will have emerged from it. […]
>
> The central question is their thought experiment, and I was intrigued to see that you speak of the "error in the last section" of my paper, but disappointed that you did not explain what my error consists in. […] I repeat that this thought experiment and its invalidity seem to me to be the central problem […] (Popper to Jauch on November 27$^{th}$, 1968. PA, 96/18)

Despite the auspices of Jauch and the willingness of Popper to seeing the note by Ramsay and Pool published, this was rejected by *Nature*. Indeed, on December 2$^{nd}$, 1968, both Ramsay and Pool received what the latter called at that time "a rather peculiar rejection letter". Whichever could have been the peculiarity of this rejection remains an unsolved historical problem, since the letter was eventually never published, not even after Popper's and Jauch's explicit insistence.[21] Only in February 1969, Pool informed Jauch about the rejection, asking for all the "suggestions and assistance" he could have been "interested in offering" (letter from Pool to Jauch on February 19$^{th}$, 1969). In fact, he continued: "both Arlan [Ramsay] and I are interested in pursuing this matter to the end. To quote Ramsay: 'It is a duty'. Popper has placed an erroneous critique of Birkhoff and von Neumann in the literature. It must not be left unchallenged".

Jauch immediately reacted against Popper, evidently believing him to be colluded with the rejection by *Nature*. He thus wrote a letter to Popper with strong accusatory tones:

> I have been informed by Professor James Pool that a comment on your unfortunate paper in Nature […] was rejected […]. You have published in a widely-read periodical criticisms of an important

---

[21] Not even the authors have today any remembrance of what the reason of that rejection was. I am thankful to Prof. Ramsay for his kind testimony (also on behalf of Prof. Pool) in two personal communications on May 15$^{th}$ and September 7$^{th}$, 2017

paper, which you have certainly misunderstood. I have tried to suggest to you to correct your mistakes yourself and that would have finished the matter. You realize of course that the entire scientific progress depends on the possibility of free exchange of scientific information and criticism. […] Did you not say yourself in the 'Open Society and its Enemies' <u>the spirit of science is criticism</u>. If you believe that, I suggest that you send the enclosed copy of the manuscript by Ramsay and Pool to Nature with your personal request that it be published" (February 24th, 1969. PA, 96/18).

Popper's indignation is here well understandable, as he immediately expressed in his reply:

> You also appear to hint in your letter […] that I have something to do with the rejection by <u>Nature</u> of Pool and Ramsay's letter. I therefore wish to state here and unambiguously that I had nothing whatever to do with it. In fact, I was waiting to hear about Pool and Ramsay's letter, and I wondered why it did not come, neither from the authors nor from yourself; nor did I hear from <u>Nature</u>.
>
> You remind me in your letter of my conviction that '<u>the spirit of science is criticism</u>'. I do not see what can give you the right to suppose that there is a need to remind me of this; or what your remark may mean unless you wish to accuse of dishonesty" (letter from Popper to Jauch on February 28th, 1969. PA, 96/18).

Within the same letter, Popper informed Jauch that he had just written a letter to the editor of Nature. Therein, Popper endorsed the publication of Ramsay and Pool's note:

> Professor Jauch asks me to support his request that this letter [Ramsay and Pool's] should be accepted for publication by <u>Nature</u>. I should indeed be glad if you could publish this letter, together with my reply (letter from Popper to Maddox on February 28th, 1969. PA, 96/18).

Reinsured, Jauch mitigated his criticism on March 3rd:

> I am really sorry that I have given the impression of suspecting you of collusion to prevent criticism to your paper. What I really wanted is to hear from you that you had nothing whatsoever to do with Nature's rejection […]. All the blame for the confusion seems thus to fall on the shoulders of Nature […]" (Jauch to Popper, PA, 96/18)

On March 17th, 1969, the editor of Nature, John Maddox, eventually sent a copy of Ramsay and Pool's critical note to Popper. Short after, Popper submitted to *Nature* a six-page note with title *Reply to Ramsay and Pool*, where he analysed one by one the three critical points put forward against his paper.[22] This reply was forwarded by Maddox to Ramsay and Pool on June 12th, 1969, and Popper was notified. The latter, however, wrote to the journal:

> Of course I do not object to your sending my reply to Ramsay and Poole [*sic*], […] but I should be anxious to get some ideas of the procedure you envisage. If Ramsay and Pool send you a reply to my reply, shall I have the opportunity to reply again? (I suppose I am to have the last word). (Popper to Maddox on June 18th, 1969. PA, 96/23).

On November 17th, Maddox sent the reply of Ramsay and Pool to Popper, asking him to "amend [his] manuscript in the light of this". However, Popper complained (December 6th, 1969):

> I do not want to go on indefinitely nor, I am sure, do you want this. […] I thought that [Ramsay and Pool's] letter and my reply would be published in Nature. Now, after reading my reply to their first letter, they have written another letter which is clearly not intended for publication, but merely for my information. Thus I do not know what they want me to publish, and what I should reply to. (letter to Maddox, PA, 96/23).

---

[22] Popper's reply also remained unpublished, but I could retrieve it in (PA, 96/23).

On December 30th, 1969, Maddox wrote to Popper the last letter of which we are aware of. Therein the editor confirmed Popper's belief that Ramsay and Pool's additional reply was not intended for publication. He then affirmed: "What I suggest is that you revise your own reply in the light of that, send it to Ramsay and Pool and then you and they can decide whether the point has now been reached at which these notes can be published." (PA, 96/23). This, unfortunately, marks the end of the correspondence between Popper and *Nature* about the point at issue, admittedly leaving still unsolved this historical case.

Another prominent physicist who was at that time working on LQM and to whom Popper sent his *Nature* paper was David Finkelstein. They met in Amsterdam at the end of 1968 and discussed Popper's critical arguments at length. Finkelstein wrote to Popper on October 16th, 1968 to point out that the major mistake of Popper was to assume that quantum logic is just any lattice for which it is possible to define an operation of orthocomplementation, but in Finkelstein's opinion quantum logic is conceptually different: it is the compound of a lattice *and* an orthocomplementation such that certain laws are satisfied. In his reply of October 29th, Popper, despite partly agreeing, stressed again that those concepts had been clarified only thanks to later contributions, whereas the original contribution of Birkhoff and von Neumann remained in his opinion flawed. Finkelstein, however, firmly replied:

> It still seems that you did not point out a single mistake in their paper. The clash you assert between their proposal and later discoveries is really a clash, as far as I can see, between hypotheses that you have appended to their paper and the facts of elementary lattice theory that I am certain they must have known at that time. (November 5th, 1968. PA, 96/13)

In the meantime, Popper was involving other physicists among his acquaintances in his enterprise. On October 17th, 1968, the Dutch theoretical physicist Hilbrand J. Groenewold commented: "I am quite happy with your criticism on Birkhoff and von Neumann […]. I have to think it over more quietly" (PA, 96/14). Also David Bohm, the great theoretical physicist who had revived the idea of hidden variables in QM, had been working himself on rebuts of arguments proposed by Jauch and his school. Popper had therefore kept him informed on his own criticisms of the LQM since the beginning of this effort (see section 3.1). On February 26th, 1969, Bohm wrote to Popper, commenting on the latter's *Nature* paper:

> My difficulties in reading anything written by von Neumann in physics is that it is deeply confused, so that even to try to understand makes my head spin widely. I have discovered that one can make no relevant criticism of a confused article. […] Your own comments 'hit the nail on the head' when you point out that 'elephants' (as well as sailing ships, sealing wax, cabbages and kings) could have been substituted for electrons in von Neumann's arguments. In other words, because von Neumann never discusses the experimental conditions, it is never determined what his symbols refer to, or whether indeed, they refer to anything at all.
>
> My advice to you is 'never entangle yourself with buzz saws, cobras, and von Neumann's articles on physics'. (PA, 96/6).

Even the Nobel Prize winner and founding father of QM Louis de Broglie wrote a short note to Popper, stating: "Yourgrau has sent me two of your articles on the interpretation of Quantum Mechanics.

I noticed with great pleasure that your ideas are very close to mine".[23] (March 4th, 1969. PA, 96/7).

On May 8th, 1969, the Argentinian philosopher and physicist Mario Bunge – who was already involved in Popper's campaign against LQM (see section 3.1 and Del Santo, 2019) and had helped him a great deal to engage with the community of physicists – discussed Popper's *Nature* paper with Scheibe, who showed a great interest (PA, 280/26). Scheibe wrote to Popper, on December 1st, 1970, levelling harsh criticisms which were at the core of the paper which was to be the first published rebut (Scheibe, 1974).

On November 30th, 1969, Popper wrote to his friend Abner Shimony –one of the protagonists of the revival of the foundations of quantum mechanics (see e.g. Freire, 2014)– trying to clarify an argument about Boolean lattices, since Shimony did not get his "point about the mathematical mistake in Jauch and Piron" at the telephone (PA, 350/7). Shimony too expressed concerns and levelled some of the criticisms that Jauch had already expressed in his letter to Popper on November 22nd, 1968, namely that the latter had made an additional assumption, i.e., the addition theorem for the probabilities.

The only rebuttal which appeared publicly a few months later was a footnote of a few lines in a publication by Jauch and Piron (Jauch and Piron, 1970). Although Popper was already aware of it, as we deduce from his letter to Shimony, he was informed of the publication of such a note directly from Jauch only on September 3rd, 1970. In the same instance, Jauch also took the opportunity to remind Popper that "it is unfortunate that [he had] still not written a correcting note". (Jauch to Popper on September 3rd, 1970. PA, 96/18). With the same letter, he also informed Popper that "A. Ramsay […] has given up hope that Nature would ever publish his and Dr. Pools [*sic*] note".

Coming back on the historical question on why Jauch had never published a rebuttal of Popper's paper, limiting himself to a heated private correspondence, one can find an answer in the final words of the mentioned letter: "I still think the correction should come from you, since this is not an interesting subject for a debate, the point at issue being entirely trivial." (PA, 96/18). Also Popper put an end to this debate that lasted more than two years, without however solving the matter in a satisfactory manner for neither of the two parties. On September 28th, 1970, he wrote the last letter to Jauch commenting the note recently appeared in (Jauch, 1970): "you read both my paper and that of Birkhoff and von Neumann in a way different from the way I read my and their papers. But I see no reason to think we should go on quarrelling about this paper: we may (for the time being) agree to disagree". (PA, 96/18).

A turning point finally happened at the beginning of the 1970s, when Popper made the acquaintance of the Canadian mathematician Simon Kochen who had already given remarkable contributions to the foundations of quantum mechanics.[24] Their interaction is particularly worth attention because this is the main reason that persuaded Popper to quit his endeavours against LQM. Popper and Kochen started corresponding about problems of quantum theory because the latter was sympathetic with Popper's realism, as he would later affirm: "my own views on quantum mechanics are somewhat closer to Einstein's and yours than to the Copenhagen interpretation" (Kochen to Popper on November 29th, 1973. PA, 315/32). On March 21st, 1974, Kochen visited Popper who was enthusiastic to have the opportunity to speak about quantum physics once more, stating: "I have had no opportunity to speak to anybody who knows any physics since I saw Shimony last in 1969". Among their topics of discussion, Popper's critique of quantum logic played a central role. Popper wrote to Kochen: "After you had left, I looked at my Birkhoff-von Neumann paper again […]. I should be deeply in your debt if you could tell me what is wrong with my simple and straightforward proof" (Popper to Kochen on March 21st, 1974. PA, 315/32). It took just a couple of days for Kochen to answer Popper's inquiry: "I have looked at your paper and I believe I see the reason for the discrepancy with Birkhoff and von Neumann's paper." (Kochen to Popper on March 23rd, 1974. (PA, 315/32). In his letter, Kochen pointed out that Popper's argument is based on a "natural misunderstanding of a misleading statement of [Birkhoff and von Neumann]". In fact, he maintained that rule (R4)[25] was not a correct translation of the original rule used by Birkhoff and von Neumann, who referred to a "dimension function" and not to a "probability measure". Moreover, in a successive letter (March 29th), Kochen also clarified that Birkhoff and von Neumann ambiguously made use of crucial concepts, using interchangeably "disjoint" and "orthogonal". This represents the resolution that Popper eventually accepted, apparently also justifying him taking the distance from his critique of LQM in later years (see footnote 4). In his reply to Kochen, Popper seems to have strongly scaled down his original motivation:

> I think I was an ass […]. I […] can maintain that Birkhoff and von Neumann were not clear about the situation, and that my criticism of their view is valid. […] But of course all of this is comparatively unimportant. The important thing is that, with the interpretation of "disjoint" as "orthogonal", your replacement of my R4 does no longer allow my proof to go through. (Popper to Kochen on March 31st, 1974. PA, 515/32).

---

[23] The original text in French, reads: "Yourgrau m'a communiqué deux articles de vous sur l'interprétation de la Mécanique Quantique. J'ai constaté avec grand plaisir que vos idées se rapprochent beaucoup des miennes.". The second paper that de Broglie referred to in the letter is here (Popper, 1967).

[24] Together with Ernst Specker, Kochen had proposed the so-called Kochen-Specker theorem, an important no-go theorem for non-contextual hidden variables (Kochen and Specker, 1967).

4.  Conclusions

In this paper, I have reconstructed the involvement of Popper in a number of activities conducted against the approach to foundations of physics, known as the *logic of quantum mechanics*. I have shown in great detail that Popper's effort went far beyond the only short published article (Popper, 1968), encompassing at least three longer manuscripts (that I partly retrieved in the archives). Moreover, I showed that the expected but apparently missing reactions to such criticism actually happened in the form of a harsh debate through private correspondence with some of the major exponents of the "school" of quantum logic. Although Popper's criticisms turned out to be mostly based on misconceptions rooted in effectively ambiguous statements of Birkhoff and von Neumann –and Popper eventually abandoned these ideas– this historical accident serves a broader scope. In fact, Popper's neglected campaign highlights how consistent was Popper's philosophy of quantum mechanics, which remained substantially unchanged (with few exceptions, such as his interpretation of quantum probabilities) for almost six decades. And one of the central elements of Popper's philosophy was the opposition against any form anti-realism, hence against Heisenberg's uncertainty principle (and Copenhagen interpretation along with it), and thus finally against LQM. At the same time, this instance helps to shed light on the period of renaissance of quantum foundations at the end of the 1960s. In fact, while this renewed interest towards fundamental and philosophical questions was long overdue, it arose in an atmosphere of fierce fights between orthodoxy (i.e., the Copenhagen interpretation) and its dissidents, which included Karl Popper.

---

[25] Kochen uses R4 in its correct form, see footnote 12.

## Acknowledgements

I am thankful to Mag.a Nicole Sager and Dr. Manfred Lube from the Karl Popper Sammlung of the AAU in Klagenfurt for granting me the access to their archives and the reproduction of Popper's original correspondence

**Appendix: An unpublished note by Ramsay and Pool**

Reproduction of the paper "Remarks on a Paper by Karl R. Popper" by A. Ramsay and J. C. T. Pool, intended for publication in *Nature* in 1968, but remained so far unpublished (see section 3.2). Retrieved from the Popper's Archive of the AAU in Klagenfurt (PA, 96/18).

---

## Remarks on a Paper by Karl R. Popper

A recent paper [1] by Karl R. Popper in Nature contains three serious errors which we feel should be corrected. It appears that Karl Popper is unaware of the literature on the subject since 1936.

In this paper he criticizes a paper [2] by G. Birkhoff and J. von Neumann, in which they put forward heuristic arguments that the calculus of experimental propositions about a quantum mechanical system should be given by an orthocomplemented modular lattice. (Orthocomplementation is the word now used to describe an operation of the type that Birkhoff and von Neumann called a complementation. In fact, the term orthocomplementation was used in the first edition of Birkhoff's well-known book on lattice theory [3].) This meaning has been accepted for 30 years and nobody except Karl Popper, to our knowledge, has thought that Birkhoff and von Neumann were implying uniqueness of complements (current terminology) by speaking of 'the "complement" of a'. The examples they refer to make this even more plain. The subspaces of a finite dimensional vector space with inner product, over a suitable division ring, form an orthocomplemented modular lattice, and Birkhoff and von Neumann were the first to explain this. This is one of the major points of interest in the paper, mathematically speaking. In fact the "re-interpretation" referred to by Popper at the top of the second column on page 683 is the clear natural reading of the paper, and the proposal they made was certainly feasible.

However, it has been observed by C. Piron [4] that modularity is evidently too strong a condition. This change in the Birkhoff-von Neumann proposal moves still farther from distributivity.

Concerning measures on the lattices, Birkhoff and von Neumann were certainly aware that a projective geometry, or the lattice of projections in a factor operator ring of type II, could have only one normalized valuation (properties D1 and D2 on page 832 or R1 - R4 of Popper). This was a result proved by von Neumann. Thus they could not have intended that the logic of a system be required to have enough of <u>them</u> to determine the order. The valuation property was suggested for an "a priori thermodynamic probability", as one conceivable physically motivated argument for modularity. On the other hand, in the first section of physical heuristics, no properties are given for the probability, in a statistical ensemble, of an experimental proposition. Consideration of the Hilbert space and operator ring examples motivates the commonly accepted property of additivity on orthogonal sequences:

(*) if $a_1, a_2, \ldots$ are orthogonal with least upper bound $a$, then
$$P(a) = \sum_{n=1}^{\infty} P(a_n).$$

This is occasionally weakened to additivity on finite orthogonal sets, which is equivalent to D2 (and R4) in the Boolean case. The Lattice of closed subspaces of a Hilbert space has enough probability measures (satisfying (*)) to determine the order, and is certainly not Boolean.

Perhaps more to the point of Popper's criticism of Birkhoff and von Neumann is their exclusive concern was with the logic. They mentioned probability in the motivating material only.

Our last point concerns Popper's discussion of the distributive law, on page 685. The experiment referred to is distinctly a quantum mechanical one. In quantum mechanics one can consider a wave function $\varphi$ such that $\varphi(x,y,-z) = \varphi(x,y,z)$ and ask the question, b, "is the wave packet $\psi$ in the state $\varphi$?" Then b is an experimental proposition, but in classical mechanics this cannot be done. The questions a, "is $\psi(x,y,z) = 0$ for almost all $z \leq 0$?" and a', "is $\psi(x,y,z) = 0$ for almost all $z \geq 0$?" can also be observed, and in fact no $\psi$ can have probability one for both b and a, or for b and a'. Thus $ba = 0$ and $ba' = 0$, which is simply to expand the valid Birkhoff-von Neumann argument